\documentclass[]{aa}
\usepackage{graphicx}
\newcommand{\be}{\begin{equation}}
\newcommand{\ee}{\end{equation}}
\begin{document}
\title{Recurrent microblazar activity in Cygnus X-1?}
\author{G. E. Romero\inst{1,2,}\thanks{Member of CONICET}, M. M. Kaufman Bernad\'o,\inst{1} and
I.F. Mirabel\inst{2,3,}$^{\star}$}
\offprints{G.E. Romero\\ \email{romero@venus.fisica.unlp.edu.ar}}
\institute{Instituto Argentino de Radioastronom\'{\i}a, C.C.5,
(1894) Villa Elisa, Buenos Aires, Argentina \and
CEA/DMS/DAPNIA/Service D'Astrophysique, Centre d'Etudes de Saclay,
F-91191 Gif-sur-Yvette, France \and Instituto de Astronom\'{\i}a y
F\'{\i}sica del Espacio/CONICET, C.C. 67, Suc. 28, Buenos Aires,
Argentina}
\date{Received / Accepted}
\abstract{Recurrent flaring events at X- and soft gamma-ray
energies have been recently reported for the galactic black hole
candidate Cygnus X-1. The observed fluxes during these transient
outbursts are far higher than what is observed in ``normal''
episodes. Here we suggest that the origin of this radiation is
non-thermal and produced by inverse Compton interactions between
relativistic electrons in the jet and external photon fields, with
a dominant contribution from the companion star field. The
recurrent and relatively rapid variability could be explained by
the precession of the jet, which results in a variable Doppler
amplification. }
\titlerunning{Microblazar activity in Cygnus X-1?}
\authorrunning{G.E. Romero et al.}
\maketitle
\section{Introduction}
Cygnus X-1 is the most extensively studied black hole candidate in
the Galaxy. It is a very bright X-ray binary with a compact object
of $\sim 10.1$ $M_{\odot}$ and a companion O9.7 Iab star of
$\sim17.8$ $M_{\odot}$ (Herrero et al. 1995), at an estimated
distance of $\sim 2$ kpc (e.g. Gierli\'nski et al. 1999 and
references therein). As in other sources of this type, the X-ray
emission switches between soft and hard states, being most of the
time in the latter. The spectrum in both states can be
approximately represented as the sum of a blackbody plus a power
law with exponential cut-off (e.g. Poutanen et al. 1997). During
the soft state the blackbody component is dominant and the power
law is steep, with a photon spectral index $\Gamma\sim 2.8$ (e.g.
Frontera et al. 2001). During the hard state more energy is in the
power law component, which is then harder, with photon indices
$\sim 1.6$ (e.g. Gierli\'nski et al. 1997).
The blackbody component is usually understood as emission from a
cold, optically thick accretion disk, whereas the power law
component is thought to be originated in an optically thin hot
corona by thermal Comptonization of disk photons (Poutanen et al.
1997, Dove et al. 1997). The hot corona would fill the inner few
tens of gravitational radii around the black hole. The accretion
disk penetrates only marginally in the coronal region. In the hard
state the thermal X-ray emission is dominated by the corona, with
typical luminosities of a few times $10^{37}$ erg s$^{-1}$. In the
soft states, the disk approaches to the black hole and then most
of the energy dissipation occurs through it (Poutanen et al. 1997;
see also Poutanen \& Coppi 1998).
Cygnus X-1 has a persistent, mildly variable, compact continuum
counterpart of flat spectrum (e.g. Pooley et al. 1999). During
many years, evidence for non-thermal radio jets in Cygnus X-1 was
lacking, despite the efforts of the observers (e.g. Mart\'{\i} et
al. 1996). Finally, the jet was detected by Stirling et al. (2001)
at milliarcsecond resolution using VLBA observations. The jet-like
feature extends up to $\sim 15$ mas with an opening angle of less
than 2 degrees. The spectum seems to be flat, and no counterjet is
observed. The total radio emission at 8.4 GHz is $\sim 11$ mJy,
with variations of $\sim2$ mJy over timescales of 2 days (Stirling
et al. 2001). The average angle with the line of sight, if the jet
is perpendicular to the disk, seems to be $\sim 30^{\circ}$
(Fender 2001).
Very recently, the Interplanetary Network detected a transient
soft-gamma ray event from the general direction of Cygnus X-1
(Golenetskii et al. 2002). Analysis of previous data indicates
that at least other two events were observed during 1995. These
latter events were also detected by BATSE instrument on the
Compton Gamma-Ray Observatory, suggesting that they were
originated in Cygnus X-1 (Schmidt 2002). The luminosities above 15
keV of the outbursts were in the range $1-2\times 10^{38}$ erg
s$^{-1}$, much higher than the typical thermal luminosity in the
hard state.
In this letter we suggest that these flaring events can be
interpreted in terms of non-thermal microblazar activity (Mirabel
\& Rodr\'{\i}guez 1999). We study the effects of the interaction
of the relativistic jet with the ambient photon fields from the
accretion disk, the corona, and the companion star, and we
calculate the expected non-thermal contribution to the keV-MeV
spectrum. The recurrent character of the events can be explained
through variable Doppler boosting originated in the precession of
the jet (Kaufman-Bernad\'o et al. 2002). In the next section we
present the model, and then we discuss the implications.
\section{Non-thermal high-energy emission for Cygnus X-1}
We shall consider the effects of the injection of a relativistic
leptonic jet at a few Schwarzschild radii from the central black
hole and its subsequent propagation through the ambient photon
fields (see a sketch of the situation in Figure 1). The individual
electrons have Lorentz factors $\gamma$ in the lab frame and the
flow is assumed to have a bulk Lorentz factor $\Gamma$. In
accordance to the disk/jet symbiosis model (e.g. Falcke \&
Biermann 1999, Markoff et al. 2001) and the estimated accretion
rate of Cygnus X-1 ($\sim 10^{-8}$ $M_{\odot}$ yr$^{-1}$, e.g.
Poutanen et al. 1997), we adopt a mean particle density
$n=10^{14}$ cm$^{-3}$ for the jet at $10 r_{\rm s}$ from the black
hole. The electron energy distribution will be assumed to be a
power law given by (in the lab frame, Georganopoulos et al. 2001):
$n(\gamma)=\frac{k}{4\pi}D^{2+p}\gamma^{-p}P(\gamma_{1}D,\gamma_{2}D,\gamma)$,
where $k$ is a constant and
$D=\left[\Gamma\left(1-\beta\cos\phi\right)\right]^{-1}$ is the
usual Doppler factor: $\beta$ and $\phi$ are the bulk velocity in
units of $c$ and the viewing angle, respectively. The function $P$
is 1 for $\gamma_{1}D<\gamma<\gamma_{2}D$ and 0 otherwise.
\\
\begin{figure}
\caption{\rm
Sketch of the general situation discussed in the paper. A
relativistic jet is injected close to the black hole in Cygnus
X-1. This jet must traverse photon fields created by the cold
accretion disk, the hot corona, and the stellar companion. Inverse
Compton up-scattering of some of these photons is unavoidable.}
\label{fig1}
\vspace{1cm}
\end{figure}
\begin{figure}
\resizebox{8cm}{!}{\includegraphics{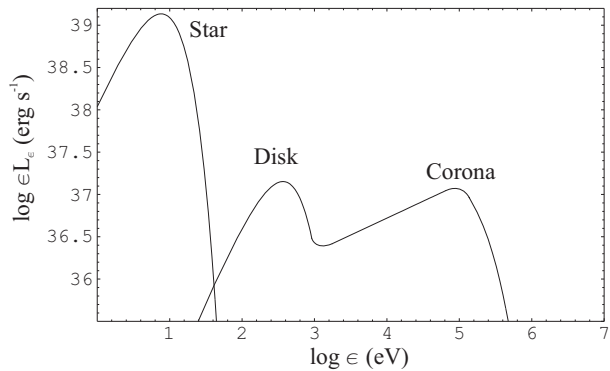}} \caption{\rm
External photon fields to which the jet is exposed.} \label{fig2}
\end{figure}
The jet will traverse first the coronal region. The parameters
that characterize this region and its photon field in the hard
state change with time. We shall assume here the typical values
given by Poutanen et al. (1997) in their Table 1. All photon
fields that interact with the jet are shown in Figure 2. The
photon field of the corona is assumed to be isotropic. External
inverse Compton interactions with these photons can be treated
using the head-on approximation in the Klein-Nishina regime since
$\gamma \epsilon_{0} \gg 1$, with $\epsilon_{0}$ the average
energy of the coronal photons in units of the electron rest energy
($\epsilon_{0} m_{\rm e} c^2\sim 100$ keV). The reader is referred
to Georganopoulos et al. (2001) for details of calculation.
In the case of the photons from the accretion disk and the
companion star we can work in the Thomson regime, taken into
account that the photons from the disc come from behind the jet,
introducing an additional factor $(1-\cos\phi)^{(p+1)/2}$ that
reduces the effects of beaming (Dermer et al. 1992). The stellar
photons, instead, can be treated as an isotropic field.
The emerging spectrum of the specific luminosity can be
approximated by a power law of index $(p-1)/2$ in the Thomson
regime (Georganopoulos et al. 2001, 2002; Kaufman-Bernad\'o et al.
2002). When $\gamma_2\gg\gamma_1$ this can be written as:
\begin{eqnarray}
    L_{\epsilon}=\frac{dL}{d{\epsilon}d\Omega}\approx D^{2+p}\frac{kV\sigma_{\rm T}c\:U2^{p-1}}{\pi
    \epsilon_{0}(1+p)(3+p)}\left(\frac{\epsilon}{\epsilon_{0}}\right)^{-(p-1)/2},
    \label{lumred}
\end{eqnarray}
where $\sigma_{\rm T}$ is the Thomson cross section, $V$ is the
interaction volume, and $U$ is the average energy density of the
photon field. In the Klein-Nishina regime, where numerical
integrations are necessary, the results significantly depart from
a power law, resulting in a softer spectrum.
The Compton losses in the different regions will modify the
injected electron spectrum, introducing a break in the power law
at the energy at which the cooling time equals the escape time.
This will occur at (e.g. Longair 1997): $ \gamma_{\rm b}=(3m_{\rm
e} c^2)/(4\sigma_{\rm T} \Gamma^2 U t_{\rm esc})$. Here $t_{\rm
esc}$ is average time spent by the particles in the field region
(typically $t_{\rm esc}\sim l/c$, with $l$ the linear size), and
the rest of the symbols have their usual meanings. The spectrum
will steepen from an index $p$ to $p+1$ for energies higher than
$\gamma_{\rm b}$. After the interaction of the jet with the
coronal region, the modified spectrum will be injected in the
stellar photon field region, suffering further losses and
modifications.
In our calculations we have adopted two different values for the
original electron energy index: a hard index $p=1.5$ and a steeper
index $p=2.3$. The jet was assumed forming an angle of 30 degrees
with the line of sight, and a bulk Lorentz factor $\Gamma=5$ was
adopted. The initial part of the jet was modeled as a cylindrical
structure; the coronal region, following Poutanen et al. (1997),
was considered as a spherical region of $\sim 500$ km in radius.
Gamma-rays produced within this region will be mostly absorbed in
the local field through pair creation. Using the simple formulae
by Herterich (1974) along with the adopted parameters for the
coronal region, we estimate an optical depth $\tau\sim 1$ for
photons of 5 MeV. The probability for a 10 MeV photon to escape
from the corona is only 0.1.
\begin{figure}
\resizebox{8cm}{!}{\includegraphics{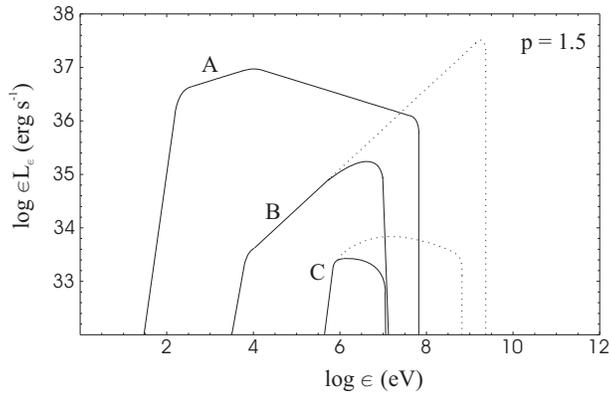}} \caption{\rm
Results of the model for an injection electron spectrum with index
$p=1.5$ in a cylindrical jet forming a viewing angle of 30
degrees. The bulk Lorentz factor is $\Gamma=5$ and the electron
power law extends from $\gamma_1=2$ to $\gamma_2=10^3$. Three
different components are shown, resulting from the up-scattering
of star (A), disk (B), and corona (C) photons. Radiation absorbed
in the local photon fields is shown in dashed lines. Notice that
the contribution from the coronal photons is not a power law
because of the Klein-Nishina effects.} \label{fig3}
\end{figure}
\begin{figure}
\resizebox{8cm}{!}{\includegraphics{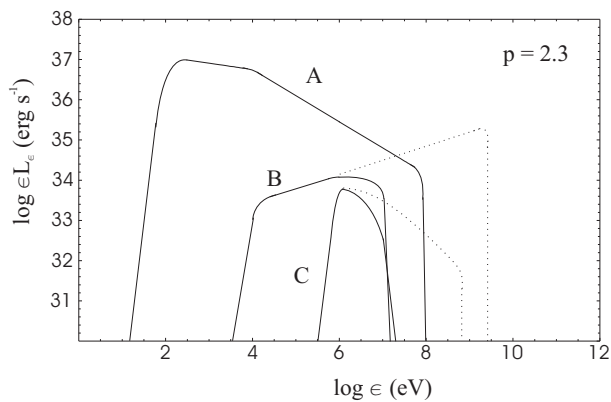}} \caption{\rm
Idem Fig. (\ref{fig3}) for $p=2.3$.} \label{fig4}
\end{figure}
The results obtained for the different models are shown in Figures
3 and 4. There, we show the different components due to external
Compton scattering of the different photon fields in the lab
frame, for the different injected electron distributions we have
assumed. This non-thermal contribution should be added to the
thermal components shown in Figure 2 in order to recover the total
emission. We can see that the non-thermal emission is dominated by
the up-scattering of the stellar photons. Above 5 MeV almost all
energy flux comes from this component.
For a viewing angle of 30 degrees, we still have the thermal
emission dominating by a factor of 5. But if we introduce the
gravitational effects of the companion star, then precession of
the disk should occur (Larwood 1998, Kaufman-Bernad\'o et al.
2002). In Figure 5 we show the modification of the  beaming
amplification factor of the external inverse Compton emission for
a precessing angle of 16.5 degrees. The time axis is normalized in
units of the period $T$. We see that there is a variation of about
1 order of magnitude in the emission measured in the observer's
frame because of the precession. This means that when the jet is
closer to the line of sight, the non-thermal luminosity can reach
values of $\sim 10^{38}$ erg s$^{-1}$, as observed in the
recurrent outbursts detected by the Interplanetary Network. The
transit through the peak of flux magnification can be very fast in
the observer's frame, leading to quick and transient states when
the total flux is dominated by the non-thermal contribution.
\begin{figure}
\resizebox{8cm}{!}{\includegraphics{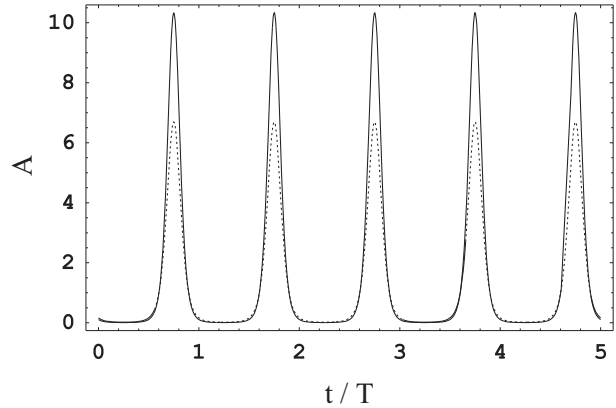}} \caption{\rm
Variation of the amplification factor for the external inverse
Compton emission from the jet (electron power-law indices of
$p=1.5$ --dots-- and $p=2.3$ --solid line--) as a function of time
in the precessing microblazar model for Cygnus X-1 (precessing
opening angle of 16.5 degrees). Time units are normalized to the
precessing period.} \label{fig5}
\end{figure}
The angular velocity of the tidally induced precession can be
approximated by $\omega_{\rm p}\approx -3Gm\cos\theta/4 r_{\rm
m}^3 \omega_{\rm d}$ (Katz 1980, Romero et al. 2000), where
$\theta$ is the half-opening angle of the precession cone, $m$ is
the mass of star, $r_{m\rm }$ is the binary separation, and
$\omega_{\rm d}$ is the Keplerian velocity of the accretion disc,
which can be written in terms of the mass $M$ of the black hole as
$\omega_{\rm d}=(GM/r_{\rm d}^3)^{1/2}$ . For $\theta=16.5$ deg as
we have assumed, the observed orbital period of 5.6 days, and a
precessing period of $\sim 140$ days (Brocksopp et al. 1999, see
below), we get a disk size of $\sim 3.9\times 10^{11}$ cm, quite
reasonable for a wind-accreting system like Cygnus X-1.
\section{Discussion}
The above outlined model incorporates the different known
components of Cygnus X-1, accretion disk, corona, stellar
companion, and relativistic jet, in an integrated picture where
transient non-thermal outbursts are a natural and expected result.
The amplitude of these outbursts can be similar to what has been
recently observed in some intriguing flaring episodes in this
source. Our model is different from the model recently proposed by
Georganopoulos et al. (2002) not only because we incorporate the
effects of precession, but also because we do not attempt to
explain the bulk of X-ray emission as non-thermal {\sl all the
time}. This emission is normally dominated by thermal
Comptonization in the hot corona around the black hole, except
during the {\sl microblazar} phase, and in this case we
incorporate the effects of the interaction of the jet with the
corona in our calculations. We emphasize that, as it is shown in
Figures 3 and 4, during the transient microblazar phase the X- and
soft $\gamma$-ray spectrum will be softer than in the normal hard
state, when the coronal emission dominates. This is an unavoidable
consequence of the steepening produced by Compton losses in the
injected electron spectrum and can be used to test our proposal,
not only through new observations of Cygnus X-1, but also of other
potential microblazars as LS5039 (Paredes et al. 2000).
Recently, Brocksopp et al. (1999) have found multiwavelength
evidence for the presence of a $142.0\pm 7.1$ days period in
Cygnus X-1. The optical and X-ray period seem to be originated in
the precession of the accretion disc (Brocksopp et al. 1999),
whereas the modulation at radio wavelengths is probably produced
by the associated precession of the jet (see Pooley et al. 1999).
The morphology of the extended radio jet, with a clear bend, is
also consistent with a precession of the inner beam (Stirling et
al. 2001). The periodic signal in the radio lightcurve, however,
is not expected to be as strong as at high energies since the
magnification factor for the synchrotron emission goes as
$D^{(3+p)/2}$, whereas for the external Compton component it goes
as $D^{2+p}$ (Georganopoulos et al. 2002).
%Additionally, only a
%fraction of the total radio flux density is originated in the
%inner precessing jet, being the rest of it produced at larger
%scales in the extended jet. Shocks are probably responsible for
%short-term radio variability and absorption in the stellar winds
%produces an additional modulation at the orbital period (Brocksopp
%et al. 2002). All this makes the analysis of radio data rather
%complex.
The time lag between the two high-energy flares observed in 1995
is $\sim 75$ days, about a half of the value reported by Brocksopp
et al. (1999), but since Cygnus X-1 is a wind-accreting system
variations in the period along a span of several years are
possible. Certainly, more observations on longer time spans are
necessary to constrain the dynamical models. In the model
presented here the duty cycle of the blazar phase is rather small,
$\sim 10\%$. Future X-ray observations of non-thermal flares can
be used for a better determination of the geometric parameters.
An interesting feature of our model is that most of the gamma-rays
produced within the coronal region will be absorbed by pair
production. Sooner or later these pairs will annihilate producing
a broad, blueshifted feature in the MeV spectrum. Details of
calculation of the emerging spectrum are beyond the scope of this
Letter (see Abraham et al. 2001), but it is clear that the
forthcoming INTEGRAL satellite will be able to probe Cygnus X-1
spectrum and its temporal evolution at this energy range, helping
to test and constrain the model here proposed. Hopefully, very
soon we will be able to clarify the role played by non-thermal
processes in this fascinating object.
\begin{acknowledgements}
G.E.R. is mainly supported by Fundaci\'on Antorchas. Additional
support comes from the research grants PICT 03-04881 (ANPCT) and
PIP 0438/98 (CONICET). He is grateful to Isabelle Grenier for
useful discussions and to the staff of the Service
D'Astrophysique, Centre d'Etudes de Saclay, where part of his
research for this project was carried out. I.F.M. acknowledges
support from grant PIP 0049/98 and Fundaci\'on Antorchas. This
research benefited from the ECOS French-Argentinian cooperation
agreement.
\end{acknowledgements}
{}

\begin{thebibliography}{}
%\bibitem{}Aharonian, F. A., \& Atoyan, A. M. 1998, New Astronomy Reviews, 42, 579
\bibitem{}Abraham, Z., Romero, G.E., \& Durouchoux, P. 2001, ESA
SP-459, 131
\bibitem{}Brocksopp, C., Fender, R. P., Larimov, V., et al. 1999, MNRAS, 309, 1063
%\bibitem{}Brocksopp, C., Fender, R. P., \& Pooley, G. G. 2002,
%MNRAS, submitted [astro-ph/0206460]
\bibitem{}Dermer, C. D., Schleickheiser, R., \& Mastichiadis, A. 1992, A\&A, 256, L27
\bibitem{}Dove, J. B., Wilms. J., Maisack, M., \& Begelman, M. G. 1997, ApJ, 487, 759
\bibitem{}Falcke, H. \& Biermann, P. L. 1999, A\&A, 342, 49
\bibitem{}Fender, R. P. 2001, MNRAS, 322, 31
\bibitem{}Frontera, F., Palazzi, E., Zdziarski, A. A., et al. 2001, ApJ, 546, 1027
\bibitem{}Georganopoulos, M., Kirk, J. G., \& Mastichiadis, A. 2001,
ApJ, 561, 111
\bibitem{}Georganopoulos, M., Aharonian, F. A., \& Kirk, J. G.
2002, A\&A 388, L25
\bibitem{}Gierli\'nski, M., Zdziarski, A. A., Done, C., et al. 1997, MNRAS 288, 958
\bibitem{}Gierli\'nski, M., Zdziarski, A. A., Poutanen, J., et al. 1999, MNRAS 309, 496
\bibitem{}Golenetskii, S., Aptekar, R., Mazets, E., et al. 2002, IAUC 7840
\bibitem{}Herrero, A., Kudritzki, R. P., Gabler, R., et al. 1995, A\&A 297, 556
\bibitem{}Herterich, K. 1974, Nat, 250, 311
\bibitem{}Katz, J. I. 1980, ApJ, 236, L127
\bibitem{}Kaufman Bernad\'o, M. M., Romero, G. E., \& Mirabel, I. F. 2002, A\&A, 385, L10
\bibitem{}Longair, M. S. 1997, High Energy Astrophysics, Cambridge
University Press, Cambridge, p.281
\bibitem{}Markoff, S., Falcke, H., \& Fender, R. P. 2001, A\&A 372, L25
\bibitem{}Mart\'{\i}, J., Rodr\'{\i}guez, L. F., Mirabel, I. F., \& Paredes, J. P. 1996, A\&A, 306, 449
\bibitem{}Mirabel, I. F., \& Rodr\'{\i}guez, L. F. 1999, ARA\&A, 37, 409
\bibitem{}Paredes, J. M., Mart\'{\i}, J., Rib\'{o}, M., \& Massi, M. 2000, Sci, 288,
2341
\bibitem{}Pooley, G. G., Fender, R. P., \& Brocksopp, C. 1999,
MNRAS, 302, L1
\bibitem{}Poutanen, J., Krolik, J. H., \& Ryde, F. 1997, MNRAS, 292, L21
\bibitem{}Poutanen, J. \& Coppi, P. 1998, Physica Scripta, T77, 57
\bibitem{}Romero, G. E., Chajet, L., Abraham, Z., \& Fan, J.H.
2000, A\&A, 360, 57
\bibitem{}Schmidt, M. 2002, IAUC 7856
\bibitem{}Stirling, A. M., Spencer, R. E., de la Force, C. J., et al. 2001, MNRAS, 327, 1273
\end{thebibliography}
\end{document}